\documentclass{edp-conf}
\usepackage{graphicx}
\begin{document}

\TitreGlobal{SF2A 2001}

\title{Thermalisation of electrons in a stellar atmosphere}
\runningtitle{Thermalisation of electrons in a stellar atmosphere}
\author{Loic Chevallier}\address{CRAL Observatoire de Lyon, Transfer group, http://www-obs.univ-lyon1.fr/\(\sim\,\)transfer/}
\maketitle
\begin{abstract}
We are interested in electrons kinetics in a stellar atmosphere to validate or invalidate the usually accepted hypothesis of thermalisation of electrons.
For this purpose, we calculate the velocity distribution function of electrons by solving the kinetic equation of these particles together with the equations of radiative transfer and statistical equilibrium.
We note that this distribution can deviate strongly from a Maxwell-Boltzmann distribution if non-LTE  effects are important.
Some results and astrophysical consequences are examined.
\end{abstract}
%
\section{Introduction}
Our work is concerned with the kinetics of electrons in a stellar atmosphere, modelled as a parallel-plane slab irradiated on a face.
Our models of atmospheres start in the deep layers of stars, where the radiative field can be described in the diffusion approximation, and end with the layers of minimal temperature, before the chromospheric raise whose effects are ignored.
The free electrons are characterized by their velocity distribution function : the electron distribution function (edf), which is calculated with the other thermodynamical quantities of the atmosphere.
Our main objective is to understand the mechanism leading to the thermalisation of electrons, where the edf tends toward the Maxwell-Boltzmann distribution.
It is accepted, in stellar atmospheres theory, that the thermalisation of electrons is effective as long as elastic collisions dominate inelastic interactions of electrons with the plasma, a rather well verified hypothesis for electrons having energies greatly below the first excitation energies of atoms and ions composing the atmosphere.
This hypothesis is not necessarily correct for faster electrons.
Our work follows the line drawn by some plasma physicists at the beginning of the 70s (Peyraud 1968, 1970; Peyraud \etal\, 1969; Oxenius 1970, 1974; Shoub 1977).
Their work demonstrated the important role played by inelastic (collisional or radiative) processes in the equilibrium reached by electrons, which can deviate considerably from the maxwellian equilibrium at high energies.
We present below a stellar atmosphere model which is not in local thermodynamical equilibrium (LTE), confirming the results anticipated in the 70s on the basis of mainly theoretical developments.
\section{The model}
This problem consists in solving the equations generally used to model a non-LTE stellar atmosphere (equation of radiative transfer, equations of statistical equilibrium, pressure equation, equation of energy, conservation of charge), coupled with the kinetic equation of electrons.
This non-linear system is difficult to solve numerically because it contains two coupled kinetic equations : one for photons, the other for electrons.
Therefore we have used the simplest model of non-LTE atmosphere : homogeneous (constant density of heavy particles \(n_0\)), isotherm (constant temperature \(T\)), and composed with hydrogen atoms with only two energy levels.
The deviation from LTE is then due to the escape of photons by the free surface.
On the other hand we have included in our model the main collision processes existing in a stellar atmosphere (elastic collisions, collisional or radiative inelastic interactions).
The elastic collision term of the kinetic equation of electrons is written in a BGK model with a velocity dependent collision frequency. This model accuratly fits the main properties of the usual Landau term (Fokker-Planck).
To solve the equation of radiative transfer, we used the codes of the transfer group in CRAL (Rutily 1992).
\section{Results}
\begin{figure}[h]
\includegraphics[width=12.5cm,height=10cm]{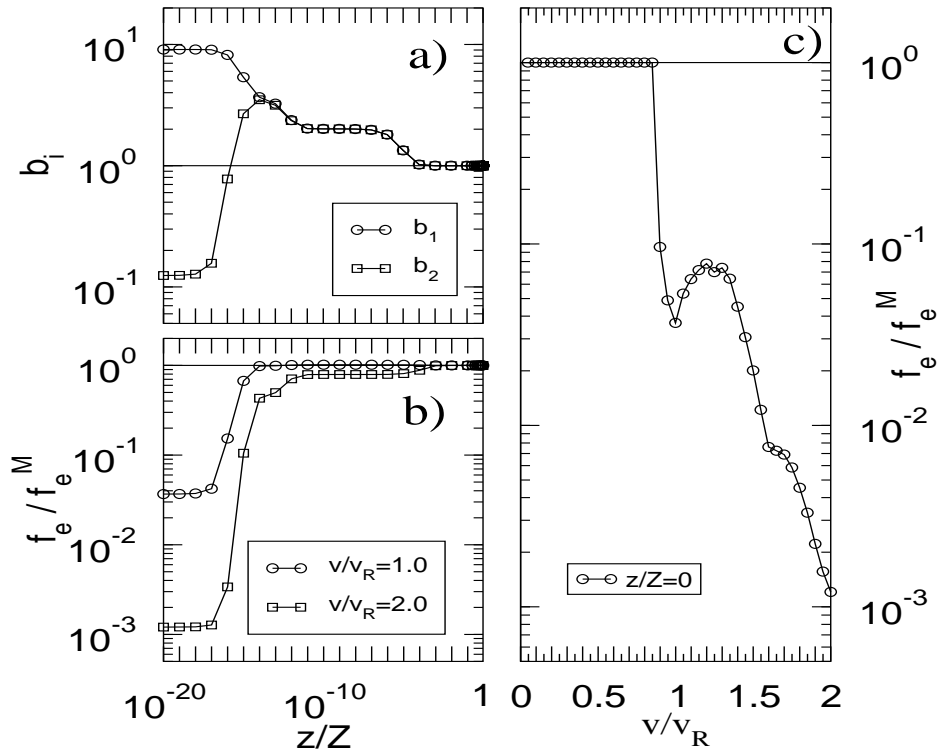}
\caption{\label{fig1}a) Deviations of the populations from their LTE values as a function of the reduced geometrical depth \(z/Z\). Coefficients \(b_i (z) = n_i ( z ) / n_i^S ( z )\), where \(n_i^S ( z )\) are the Saha densities of the hydrogen atom in energy states \(i = 1, 2\), are used to characterize non-LTE regions (\(b_i (z) \ne 1\)). b) Deviation of the edf to the maxwellian distribution \(f_e^M(v)\) as a function of the reduced geometrical depth \(z/Z\). Both curves are drawn for a given velocity \(v/v_{\mathrm{R}} = 1, 2\), where \(v_{\mathrm{R}}\) is the velocity corresponding to the Rydberg energy \(E_{\mathrm{R}} = 13.6 \,\mathrm{eV}\). c)  Deviation of the edf to the maxwellian distribution as a function of the electronic velocity \(v\), at the surface of the atmosphere \(z = 0\).}
\end{figure}

In our model, we choosed \(n_0 = 1.2 \,\mathrm{x}\, 10 ^ {23} \, \mathrm{m}^{3}\) and \(T = 6500 \,\mathrm{K}\), which are typical values in the solar photosphere.
The plasma is optically thick at all frequencies (optical thickness greater than 100), leading to a high geometrical thickness \(Z\) since there is no temperature or heavy particles density gradient.
Finally the atmosphere is irradiated on its internal boundary layer by a Planck radiation of temperature \(T\).
\newline
The figure~\ref{fig1}a is a classical diagram showing the superficial regions where the non-LTE effects are important.
Figure~\ref{fig1}b shows that the edf is not a Maxwell-Boltzmann distribution in the non-LTE region of the atmosphere (see \(v/v_{\mathrm{R}} = 2 \)), the deviation from a maxwellian distribution being important very close to the surface (\(z/Z < 10^{-15}\), corresponding to an optical depth \(\tau < 30\) in the Ly\(\alpha\) spectral line).
In figure~\ref{fig1}c, we drawn the superficial edf at \(z = 0\) as a function of the electronic velocity.
The edf tail of fast electrons is strongly depleted when electron energies are greater than the minimum excitation energy of the hydrogen atom (\(E = 10.2 \,\mathrm{eV}\), \(E / E_{\mathrm{R}} = 0.75\), \(v / v_{\mathrm{R}} = 0.866\)).
The edf tail shows successive platforms centered on \(v / v_{\mathrm{R}} = 1.25\) and \(v / v_{\mathrm {R}} = 1.65\).
\newline
These features were already described by the authors at the origin of this work, referenced at the beginning of this article.
The mechanism responsible for this effect is very well explained in Oxenius's monograph (1986), where the author outlines an interesting \emph{feedback effect} tending to amplify the deviation of the edf from a maxwellian distribution.
This mechanism starts when elastic and inelastic collision frequencies become comparable at high electronic velocities, which is the case for a weak ionization degree. 
\section{Astrophysical applications and conclusions\label{sec_appastro}}
Astrophysical consequences of this work are numerous.
In general the deviation of the edf from a maxwellian distribution has a direct effect on all thermodynamic quantities involving the edf, \emph{e.g.} collisional transition rates or spectral lines profiles. It has an indirect effect on all other characteristics of the atmosphere, because of the coupling of all equations.
Transition rates are used to solve the equations of statistical equilibrium, which lead to the populations and ionization degree of the atmosphere.
Inversion techniques of spectral lines observed by spectroscopy are also sensitive to the edf shape, so that temperatures or densities calculated with these techniques are \apriori\, affected by deviation of the edf from a maxwellian distribution (\cf \, Shoub 1983, Owocki \etal\, 1983, Salzmann \etal\, 1995).
Finally the non thermodynamical equilibrium of electrons may be at the origin of physical processes which are still not very well understood at present, for example the heating of the sun corona (\cf \, Scudder 1992,1994).
\newline
The results presented in this article are based on a very simple atmosphere model, which guarants a numerically stable solution.
In non-LTE regions close to the surface, the edf shows important deviations from the maxwellian distribution in the fast electrons tail.
Our model confirm, by means of numerical codes accurate enough to handle this complex problem, most of the physical ideas advanced thirty years ago.
Our main contribution consists in the construction of a selfconsistent model of a stellar atmosphere with \apriori\, non thermalized electrons. Also we have used very accurate radiative transfer codes.
It remains to make this model more realistic for comparison with observations.

\end{document}